\documentclass{article}
 \usepackage{amstext}
\usepackage{graphicx}

\usepackage[utf8]{inputenc} % allow utf-8 input
\usepackage[T1]{fontenc}    % use 8-bit T1 fonts
\usepackage{hyperref}       % hyperlinks
\usepackage{url}            % simple URL typesetting
\usepackage{booktabs}       % professional-quality tables
\usepackage{amsfonts}       % blackboard math symbols
\usepackage{nicefrac}       % compact symbols for 1/2, etc.
\usepackage{microtype}      % microtypography
\usepackage[preprint]{neurips_2020}
 \usepackage{amsmath,amsthm,epsfig,amsfonts,amssymb}
 \usepackage{amstext}
 \usepackage{color}
 \usepackage{float}
 \usepackage{natbib}
 \usepackage{appendix}
 \usepackage{caption}
\usepackage{subcaption}
\theoremstyle{definition}

\title{Anomaly and Fraud Detection in Credit Card Transactions Using the ARIMA Model }

\author{%
 Giulia Moschini \\
 University of Applied Sciences and Arts, Western Switzerland (HES-SO)\\  
 School of Engineering and Business Vaud (HEIG-Vd)\\
  CH-1401 Yverdon-les-Bains \\
   \And
  Régis Houssou \\
  University of Applied Sciences and Arts, Western Switzerland (HES-SO)\\  
  School of Engineering and Business Vaud (HEIG-Vd)\\
   CH-1401 Yverdon-les-Bains \\
   \texttt{regis.houssou@heig-vd.ch} \\
   \AND
  Jérôme Bovay \\
  NetGuardians SA\\
  CH-1401 Yverdon-les-Bains \\
  \texttt{jerome.bovay@netguardians.ch} \\
   \And
  Stephan Robert-Nicoud \\
  University of Applied Sciences and Arts, Western Switzerland (HES-SO)\\  
  School of Engineering and Business Vaud (HEIG-Vd)\\
   CH-1401 Yverdon-les-Bains \\
 \texttt{stephan.robert@heig-vd.ch} \\ 
}

\begin{document}

\maketitle

\begin{abstract}
This paper addresses the problem of unsupervised approach of credit card fraud detection in unbalanced dataset using the ARIMA model. The ARIMA model is fitted on the regular spending behaviour of the customer and is used to detect fraud if some deviations or discrepancies appear. Our model is applied to credit card datasets and is compared to $4$ anomaly detection approaches such as K-Means, Box-Plot, Local Outlier Factor and Isolation Forest. The results show that the ARIMA model presents a better detecting power than the benchmark models.
\end{abstract}

\section{Introduction}

In the recent years, there has been a dramatic increase in the use of credit cards as a means of payment due to their ease of use and convenience. As a response to this phenomenon, fraudsters are also adapting their malicious activities to take advantage of the situation. The extent of this issue is significant, according to the Fifth report on card fraud published by the European Central Bank [1], the total value of fraudulent transactions in the SEPA area in $2016$ amounted to $1.8$ billion euros. According to the Nilson Report, a publication covering global payment systems, the total loss due to frauds in 2018 amounted to \$27.85 billions, and it is projected to reach \$35.67 billions in $2023$ [2].

More specifically, a transaction is said to be fraudulent when it is committed by an unauthorised party and without the rightful owner and/or relevant institution knowing [3]. In these cases,  fraudsters could use the card for their personal interests depleting its resources or until they are caught or the card is blocked. This issue has sparked the interest of the both academia and industry, that are working to find solutions to this problem and to keep up with the ever-changing approaches adopted by malicious players [4]. Credit card fraud detection is now an active field of research, and it particularly hinges on the concept of automation; it is in fact not always feasible nor possible to manually review each transaction in order to establish its nature [5]. In addition to this, it is also important to consider that there is another significant human component that could make or break the attempt of a fraudster to successfully exploit a card: the promptness of the cardholders in reporting a stolen, lost or suspiciously used card [5]. This requires the implementation of automated tools for a smarter and faster detection of frauds, which has resulted in machine learning techniques being increasingly tested and implemented [6].

Various popular algorithms have been tested in this context, such as Random Forest, Logistic Regression, Decision Trees, Support Vector Machines (SVM), and Neural Networks [7], [8], [9]. Khare and Sait in [7] compare Logistic Regression, SVM, Decision Tree and Random Forest using the Kaggle dataset for credit cards containing $284,807$ transactions, $492$ of which are fraudulent. The features of the dataset are obtained using Principal Component Analysis (PCA) on the original data for confidentiality issues. The authors also state that they use the behavioural characteristics of the owner of the card, which is shown by a variable representing the spending habits of the customer as well as the month, hour of the day, geographical location and type of merchant. Experimental results show that Random Forest is the most performing algorithm, with an accuracy score of $98.6\%$, compared to the $97.7\%$ of Logistic Regression, $97.5\%$ of SVM and $95.5\%$ of Decision Tree. Varmedja  et al in [8] compare the performances of Logistic Regression, Naive Bayes, Random Forest and Multi-Layer Perceptron on the Kaggle Dataset. 

The number of features is reduced through the application of feature selection and the class imbalance addressed by oversampling with SMOTE. Their results show that Random Forest is again the best algorithm , with Accuracy, Precision and Recall equal to $99.06\%$, $96.38\%$ and $81.63\%$ respectively. Roy et al in [9] use a deep learning approach to detect frauds in credit card transactions. The dataset used in the study was provided by a financial institution and contains almost 80 million anonymised transactions performed over a period of 8 months. The authors perform feature engineering to apply field knowledge to the problem and add extra features to the original ones. 
%These include the frequency of transaction per month, dummy variables for missing data and to indicate specific merchants (i.e. the gas station is a typical location for fraudsters to test a stolen card), account history variables (i.e. number of transactions in an 8 months period). 
Due to the unbalanced nature of the dataset, the authors also perform under sampling at the account level for each unique account ID. Artificial Neural Networks (ANN), Recurrent Neural Networks (RNN) Long Short-term Memory (LSTM) and Gated Recurrent Unit (GRU) are compared in this study; the results highlight that GRU presents the best performance with an accuracy score equal to 91.6\%, followed by 91.2\% (LSTM), 90.4\% (RNN) and 88.9\% (ANN). 

As can be noted, there is a common fundamental issue in these approaches: the unbalanced nature of the datasets. In the context of credit card, fraud detection is in fact expected that the dataset will be very unbalanced, which greatly hinders the performance of supervised learning techniques [6]. Another issue involves the lack of properly labelled data, which again represents a substantial obstacle. Finally, many models lack the adaptability required to take into account the fact that the spending behaviour of customers is likely to change over time [6]. In order to tackle these problems, we propose a model that does not require the knowledge of ground truths and that is designed to make the spending behaviour of the customer as the main source of information when categorising transactions as either legitimate or fraudulent. More specifically, we frame the problem as an anomaly detection task in time series, where the variable represented by the time series is the daily count of transactions for a given customer. We propose a method making use of the ARIMA model and of a rolling windows approach to flag suspicious number of transactions as anomalies, which will be discussed in-depth in the following sections.

\section{Fraud Detection with Time Series Approach}
\subsection{ARIMA model with Time Series Analysis}
%The term time series refers to a "sequential set of data points, measured typically over successive times" [10, p.12]. The records could either be univariate or multivariate, depending on the number of variables considered, as well as discrete or continuous. A time series comprises of four fundamental components: 
%\begin{enumerate}
%\item \textit{Trend}, which indicates the general tendency of the time series to increase or decrease. 
%\item \textit{Seasonal}, which refers to the seasonal fluctuations that might occur every month or quarter. 
%\item \textit{Cyclical}, which indicates changes in response to cyclical events that usually occur every two or more years. 
%\item \textit{Irregular}, which refers to the unpredictable variations that happen without a particular pattern. 
%\end{enumerate}
%We can either assume that these components affect each other or that they are independent; in the first case, the time series can be described by the multiplicative model (1), whereas in the latter, the time series is represented by the additive model (2). 
%\begin{eqnarray}
%Y(t) = T(t) * S(t) * C(t) * I(t)\\
%Y(t) = T(t) + S(t) + C(t) + I(t)\
%\end{eqnarray}
%Where $Y(t)$ is the observation and $T(t)$, $S(t)$, $C(t)$, and $I(t)$ are the trend, seasonal, cyclical and irregular components respectively. \\

Two widely used models for time series are the \textit{Autoregressive} (AR) and the \textit{Moving Average} (MA) models, which can be used together as an \textit{Autoregressive Moving Average} (ARMA) model. ARMA(\textit{p}, \textit{q}) is the combination of the AR(\textit{p}) and MA(\textit{q}) models, and can be used with univariate time series. \\

\begin{itemize}
\item \textit{Autoregressive Model}

The AR(\textit{p}) model is defined by the equation below; it assumes that there is a 	dependent linear relation between the observation and the values of a specified number of lagged (previous) observations plus an error term. 
\begin{eqnarray}
X_t = c + \sum_{i=1}^{p}\phi_i X_{t-i} + \omega_t   
\end{eqnarray}
Where $\phi = (\phi_1, \phi_2,..., \phi_n)$ are the coefficients of the model, $p$ is a non-negative integer, $c$ is a constant and $\omega_t \sim N(0,\sigma^2)$.
	
\item \textit{Moving Average Model}

The MA(\textit{p}), model is defined by the equation below; it makes use of the dependency 			between an observation and the residual errors resulting from the application of a moving 			average model to lagged observations. 
\begin{eqnarray}
X_t = \mu + \sum_{j=1}^{q}\theta_j \omega_{t-j} + \omega_t
\end{eqnarray}
Where $\mu$ is the mean of the series, $\theta = (\theta_1, \theta_2,..., \theta_n)$ are the coefficients of the model, $q$ is the order and $\omega_t \sim N(0,\sigma^2)$.
\end{itemize}
The ARMA model, resulting from the combination of these two models, is defined as follows:
\begin{eqnarray}
X_t = c + \omega_t + \sum_{i=1}^{p}\phi_i X_{t-i} + \sum_{j=1}^{q}\theta_j \omega_{t-j}
\end{eqnarray}
Where $p$ refers to the order of the AR model and $q$ refers to the order of the MA model. The main assumption in time series analysis is that the time series is \textit{stationary}, meaning that its mean and variance are constant over time; however, this is not the case in many practical situations [10]. The solution to this can be found in the generalisation of the ARMA model: the \textit{Autoregressive Integrated Moving Average}(ARIMA) model. ARIMA introduces the possibility to apply differencing to the data points of time series in order to make it stationary [10]. ARIMA is now one of the most popular, flexible and simple models to fit a time series 	[10]; it is defined as ARIMA(\textit{p}, \textit{d}, \textit{q}) where \textit{p} and \textit{q} represent the orders of the AR and MA models and the \textit{d} indicates the degree of differencing. In the context of fraud detection, time series can be used as a tool when working with \textit{aggregated features}. Aggregation is often used to derive new features from the original ones in order to feed to the model some information that is thought of and expected to be more relevant than the features per se. The number of daily transactions or the total amount spent in a week are examples of aggregated features [5].

\subsection{Estimation Process of ARIMA}
When using ARIMA, care should be taken to identify the combination of parameters that best represents the data; Box-Jenkins is a method proposed by George Box and Gwilym Jenkins in [11] that is frequently used when tuning an ARIMA model. The method is composed of three steps:
\begin{enumerate}
\item \textit{Identification}, which refers to the use of all available data and related information to select the model that best represents the time series. This phase should however be split into two sub-steps: 
\begin{enumerate}
\item Differencing \\
The first step requires to establish whether the time series is stationary or not to determine whether it requires differencing. The Augmented Dickey-Fuller (ADF) test is a technique that can be used to verify if the time series on hand is stationary. 
%This is a type of statistical test called a unit root test; the intuition behind it is to determine whether a time series is strongly related to a given trend. 
The null hypothesis of the ADF test states that the time series can be represented by a unit root, meaning it presents a time-dependent structure and that is, thus, not stationary; consequently, rejecting the null hypothesis implies that the time series is stationary. 
\item Configuration of \textit{p} and \textit{q} \\
During this phase, it is helpful to use the correlogram to visualize the autocorrelation function (ACF) and the partial autocorrelation function (PACF) that can help to determine a suitable choice for the orders $p$ and $q$.  The fundamental difference between the two functions is that the PACF removes the linear dependence between the intermediate variables in order to return only the correlation between the present and lagged value. Briefly, whereas the autocorrelations function of AR(p) tails off, its partial autocorrelation function has the cutoff after the lag $p$. Conversely, the autocorrelations function of MA(q) has a cutoff after the lag $q$ while its its partial autocorrelation function tails off. 

%The plots show on the y-axis the autocorrelation values (from -1 to 1) and on the x-axis the size of the lags between the elements of the time series. The ACF plot can be used to estimate the \textit{q} parameter, that is, the order of the MA model. The \textit{p} parameter (AR model) can be estimated looking at the PACF plot, which shows the direct correlation between an observation and the ones at previous time steps. The fundamental difference between the two plots is that the PACF removes the linear dependence between the intermediate variables in order to return only the correlation between the present and lagged value. The ACF and PACF can be plotted together with the 95\% confidence intervals as parallel lines to the x-axis in order to identify the bars that cross them (i.e. significant points).
\end{enumerate}
\item \textit{Estimation}, which refers to the training phase. Once the values of \textit{p}, \textit{d}, \textit{q} have been established, the $\phi$ and $\theta$ coefficients can be estimated. This method uses the maximum likelihood estimation process, which is solved by non-linear function maximisation; for more details about this phase the reader is referred to [11], [12]. 
\item \textit{Diagnostics}, which refers to the evaluation of the model and identification of improvements. This step involves the determination of issues in the model to verify whether it is able to effectively summarise the underlying data. The forecast residuals provide an important source of information for diagnostics. In an ideal model, the error will resemble white noise and will be normally distributed with a mean of 0 and a symmetrical variance. In addition to this, an ideal model would also leave no temporal structure in the residuals, as they should have been learned.
\end{enumerate}

\subsection{Fraud Detection with ARIMA Model on Daily Counts of Transactions}
\begin{figure}[H]
\begin{center}
\includegraphics[width=5in]{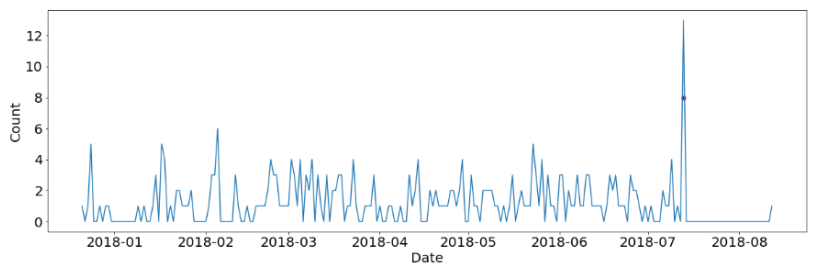}
\end{center}
\caption{Plot of daily number of transactions for a customer in the dataset. Legitimate transactions are in blue dot whereas Fraudulent transactions are red dot}
\label{fig1}
\end{figure}

Our idea is to use ARIMA on time series representing the daily count of transactions for a given customer to detect frauds. This is based on an important point: we assume that the number of daily transactions for a given customer follows a certain pattern [13]. On a high level, the task of fraud detection in this context is based on the assumption that it is possible to recognise, and hence model, the regular spending behaviour of the customer; once this has been learned, any discrepancies and deviations from it would be likely to be frauds. We can also refer to such deviations as \textit{anomalies}. An anomaly is a point in a dataset whose characteristics are significantly different compared to the other points; building from this, anomaly detection is the process to isolate such points by determining when they are deviating from the expected behaviour [14]. ARIMA will be used to try and model the legitimate spending behaviour of the customer and to produce a forecast. The intuition behind this setting can be easily explained graphically. Figure \ref{fig1} shows the daily transactions of a credit card for a customer chosen in our dataset; more details about this dataset will be given in the next section. The number of legitimate transactions happening each day for such customer are in blue dot, whereas the number of frauds are in red dot. A significant peak is observed at the same day of fraudulent transactions.  Based on this information, it could be argued that an anomaly detection approach based on the identification of anomalous counts of daily transactions may lead to the detection of frauds. In order to detect frauds, the following steps are proposed:

\begin{enumerate}
\item The time series is split into training and testing set; it is important that the training set only contains legitimate transactions so that the model would learn the legitimate behaviour of the customers. This should then allow for the identification of anomalies. 
 \item In the training set, based on the legitimate transactions the order of the ARIMA model is identified using the Box-Jenkins method and then the parameters of ARIMA are estimated. During this phase, care is taken to ensure that the estimated coefficients are significant and that there is no temporal structure left in the residuals. Finally in the testing set, the one-step ahead prediction is performed by rolling windows.
\item In order to detect fraud in the testing set, the errors are calculated in terms of difference between the predicted and actual daily count of transactions. Then, the Z-Scores are computed and used to flag the anomalies (i.e. the frauds). The Z-Score is calculated as 
\begin{eqnarray}
\text{z-score} = \frac{x-\mu}{\sigma}
\end{eqnarray}
where x is the prediction error on the daily count of transaction in the testing set. $\mu$ and $\sigma$ are the mean and the variance based on the errors of In-Sample prediction based on the training set using our model. If the Z-Score is greater than a threshold, the day is flagged as anomalous (i.e. as fraud). 
\end{enumerate}

\begin{table}
\caption{Frequency of fraud and legitimate transactions in the whole dataset\vspace{2mm}}
\label{table:1}
\centering
\begin{tabular}{c|c|c|c}
 \toprule
%\hline
& Legitimate & Fraud & Total\\ 
  \midrule
Number & 11384 & 87 & 11471 \\ 
Percentage & 99.24\% & 0.76\% & 100\% \\ 
 \bottomrule
\end{tabular}
\end{table}

\section{Application to Dataset}
\subsection{Dataset Description}
\begin{figure}[H]
\begin{center}
\includegraphics[width=5in]{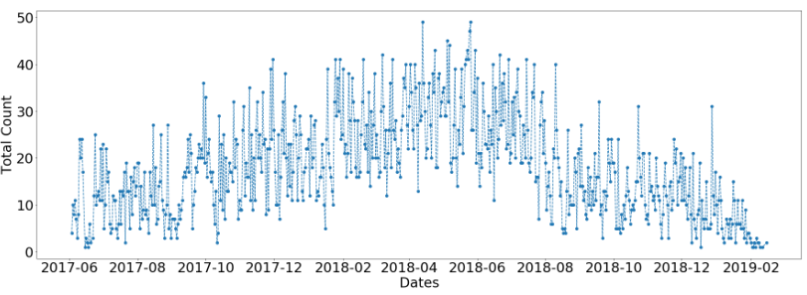}
\end{center}
\caption{Number of daily transactions summing up all customers}
\label{fig2}
\end{figure}

The dataset used for this study was provided by NetGuardians SA and contains information about credit card transactions for $24$ customers of a financial institution; it covers the period from June $2017$ to February $2019$. For reasons of confidentiality, the name of the financial institution will not be mentionned. Each row is related to a customer ID and represents a transaction with its various features (i.e. timestamp, amount..) including the class label (1 for fraud and 0 for legitimate transaction). An important aspect is that each of the 24 customers presents at least one fraud in the whole period. Figure \ref{fig2} and table \ref{table:1} show the number of daily transactions for all customers and the frequency of fraud and legitimate transactions in the whole dataset. We remark that the dataset is highly imbalanced with a proportion of fraud of $0.76\%$.

\begin{figure}[H]
\begin{center}
\includegraphics[width=5in]{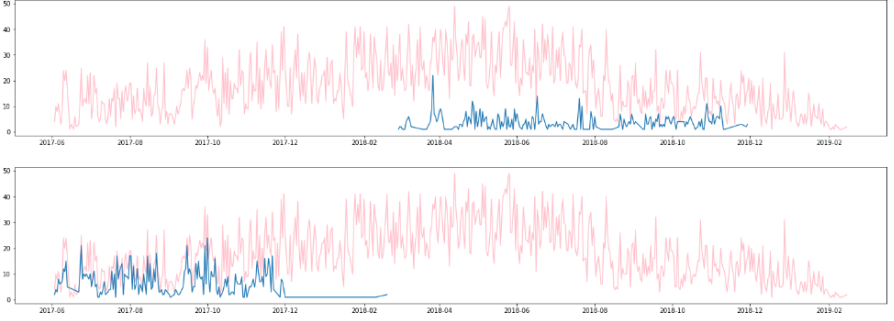}
\end{center}
\caption{Number of daily transactions: The blue dot represents a specific customer and the red dot represents all customers}
\label{fig3}
\end{figure}

\begin{figure}[H]
\begin{center}
\includegraphics[width=5in]{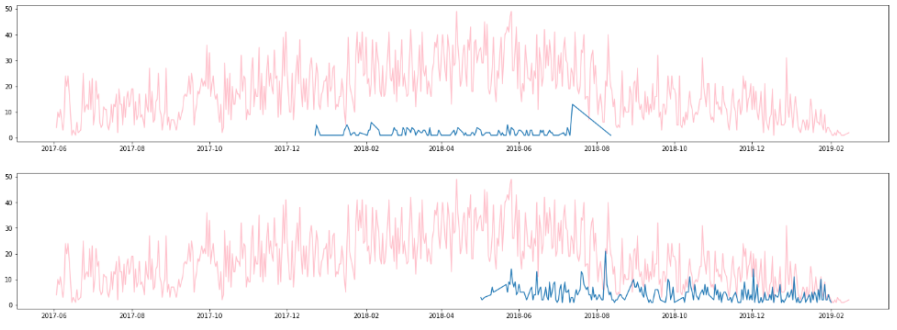}
\end{center}
\caption{Number of daily transactions: The blue dot represents a specific customer and the red dot represents all customers}
\label{fig4}
\end{figure}

However, it is important to notify that the customers are not necessarily active during the whole period. In fact, as illustrated in Figure \ref{fig3} and Figure \ref{fig4}, some of them perform transactions only in the first part of the considered time frame, others only at the end, and others in the middle. Our approach based on ARIMA model requires in the training set a sufficient legitimate transactions in order to learn the legitimate behaviour of the customers. In addition, our approach  requires at least one fraud in the testing set to evaluate the performance of the model. In this context, initially we propose to split the dataset into the training and testing set with 70-30 ratio. With this setting, there is at least one fraud in the testing set and no fraudulent transactions in the training set but unfortunately this reduces  the number of customer's time series from $24$ to $9$. Table \ref{table:2} summarises the composition of the final 9 time series that will  be used in the next section. The last column indicates the number of frauds over the total number of transactions happening in the same day; as can be seen, only in one of the time series (number $10$) frauds happen in two different days. 

\begin{table}[h]
\caption{Structure of the $9$ time series\vspace{2mm}}
\label{table:2}
\centering
\begin{tabular}{c|c|c|c}
 \toprule
Time Series ID & \# days in train & \# days in test & fraud proportion\\ 
\midrule
0 & 192 & 83 & 1/14 \\ \hline
4 & 193 & 84 & 1/3 \\ \hline
5 & 192 & 83 & 1/16 \\ \hline
7 & 186 & 80 & 1/11 \\ \hline
8 & 131 & 57 & 3/15 \\ \hline
9 & 164 & 71 & 8/21 \\ \hline
10 & 193 & 84 & 4/17 \\ \hline
15 & 191 & 82 & 1/11 and 1/2 \\ \hline
17 & 119 & 51 & 2/12 \\ 
\bottomrule
\end{tabular}
% \vspace{3mm} 
%\caption{Structure of the $9$ time series}
\end{table}
 
\subsection{Application of ARIMA Model for Daily Counts of Transactions}

The previously outlined steps are performed for each of the $9$ time series separately. These are now described in detail for just one of the time series for the sake of clarity and brevity as an illustration. As already discussed, the first step involves establishing whether the time series is stationary. To do this, we perform the ADF test whose results are shown in the table \ref{table:3}. 

\begin{table}[htb]
\caption{Statistics of ADF on the stationarity for one time series\vspace{2mm}}
\label{table:3}
\centering
\begin{tabular}{c|c}
\toprule
t-Statistic & -8.73162539099 \\ \hline 
P-value & 3.18017662959e-14 \\  
\bottomrule
\end{tabular}
% \vspace{3mm} 
%\caption{Statistics of ADF on the stationarity for one time series}
\end{table}

\begin{figure}%[htb]
\centering
\includegraphics[scale=0.35]{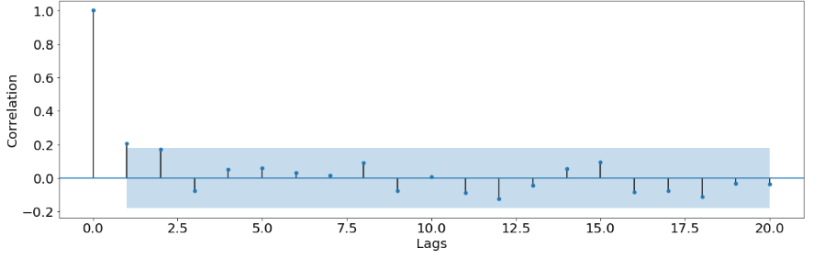}
\includegraphics[scale=0.35]{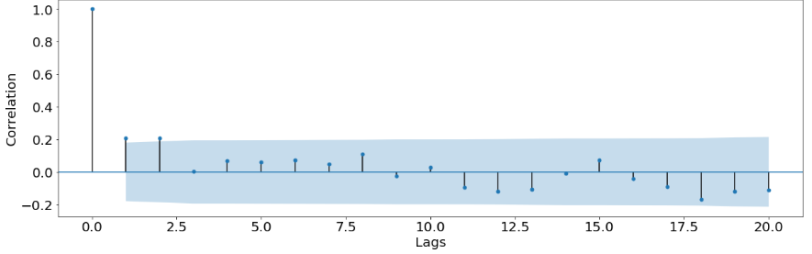}
\includegraphics[scale=0.35]{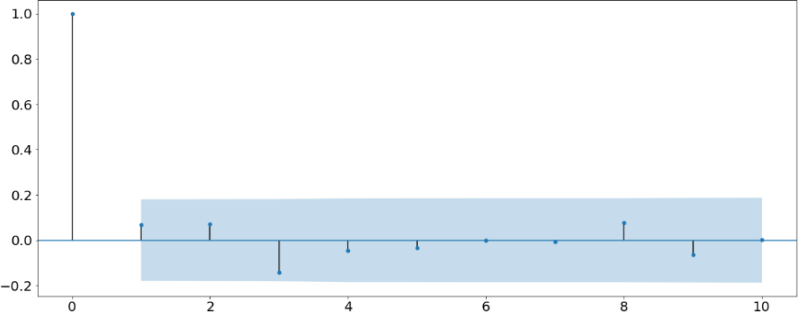}
\caption{(a) Partial Autocorrelation plot for sample time series, (b) Autocorrelation plot for sample time series, (c) Correlogram of residuals}
\label{fig5}
\end{figure}

It shows that the time series is stationary with significant result. Next, Figures \ref{fig5}(a) and \ref{fig5}(b) show the PACF and ACF that are used to determine the best values for the order \textit{p} and \textit{q} of the ARIMA model. For this time series, there may be a drop-off in the PACF at lag $1$ and  in  the  ACF  at  either  lag  $1$ or $2$ suggesting an ARIMA(1,0,1) or ARIMA(1,0,2). The steps for the parameters estimation and the residuals analysis in the training set conduct to select among the two models, the model ARIMA(1,0,2) as a good model for this time series and will be used to make forecasting. Figure \ref{fig5}(c) shows the correlogram of the residuals for the selected model and this confirms that they have a white noise pattern.These above steps are performed for all the other 8 time series; in some cases, the configuration may require multiple attempts to find the best parameters. All parameters passed on to the next stage of the study are found to be significant. It is important to mention that for the forecasting in the testing set, we set the threshold to $3$. So, when the Z-Score is greater than 3, there is fraud.

%\begin{figure}[htb]
%\centering
%\includegraphics[scale=0.6]{ARIMA_step2.png}
%\caption{Partial Autocorrelation plot for sample time series}
%\label{fig5}
%\end{figure}

%\begin{figure}[htb]
%\centering
%\includegraphics[scale=0.6]{ARIMA_step3.png}
%\caption{Autocorrelation plot for sample time series}
%\label{fig6}
%\end{figure} 

%\begin{figure}[t]
%\centering
%\includegraphics[scale=0.6]{ARIMA_diagnostics_correlogram.png}
%\caption{Correlogram of residuals}
%\label{fig7}
%\end{figure}

%\begin{figure}[!ht]\centering
%\subfloat[ Partial Autocorrelation]{\includegraphics[height=0.6in]{ARIMA_step2.png}\label{fig5}}
%\subfloat[Autocorrelation ]{\includegraphics[height=0.6in]{ARIMA_step3.png}\label{fig6}}
%\subfloat[Correlogram of residuals]{\includegraphics[height=0.6in]{ARIMA_diagnostics_correlogram.png}\label{fig7}}
%\caption{Partial autocorrelation and autocorrelation plots for sample time series and correlogram of the residuals for the selected model}
%\end{figure}

\subsection{Benchmark Models}

Our model is compared to $4$ different models of anomaly detection such as the Box-Plot, the Local Outlier Factor (LOF), Isolation Forest and the K-means. Each benchmark model is briefly explained in the following section.  
%There are many models that can be used to detect anomalies, some of the most widely used ones are now discussed and will be used as a benchmark to compare the performance of our model in the following sections. 
\paragraph{Box-Plot}
Box-Plots are used in the context of exploratory data analysis; they can be used to graphically represent data using their descriptive statistics.  
%\begin{itemize}
%\item The \textbf{minimum}, that it the lowest data point excluding the outliers. 
%\item The \textbf{maximum}, that is the largest data point excluding the outliers.
%\item The \textbf{median} (or second quartile), that is the middle value of the dataset. 
%\item The \textbf{first quartile}, which is the middle value between the smallest number and the median of the dataset.
%\item The \textbf{third quartile}, which is the middle value between the largest number and the median of the dataset. 
%\end{itemize}
Box plots do not make any assumptions about the statistical distribution followed by the sample, meaning that potential outliers are identifies solely based on the degree of dispersion of the data points in the sample. Box-Plots are very useful because they can be used to effectively identify patterns in groups of numbers that might be invisible to the human eye [15]. Being a visual tool, box plots are often used to increase our understanding of data allowing for a better interpretation for quantitative data [15]. We apply Box-Plot on the entire dataset (for each time series); however, only the testing portion of the dataset is considered to calculate the results. This is done for consistency reasons in order to ensure a fair comparison of the performances.

\paragraph{Local Outlier Factor (LOF)}
Local Outlier Factor (LOF) is an algorithm introduced by Breunig, Kriegel, T. Ng and Sander in 2000 that is aimed at the identification of anomalous data points based on their local deviation from their neighbours. LOF is a density-based algorithm, and it is centred on the concept of \textit{degree} of being an outlier [16], as opposed to a binary classification of outliers. The model is local because the anomaly score assigned to each point derives from the degree of isolation of that point compared to the its $k$ neighbours, where $k$ can be specified. More precisely, the locality of a point is given by its k-nearest neighbours. A point is considered to be an outlier when its local density results to be significantly lower than the densities of its neighbours [17]. For more details about LOF, see [16]. As for Box-Plot, LOF is applied on the entire dataset only considering the testing set to calculate the results. As already explained, this is done in order to retain consistency across the tests. 
  
\paragraph{Isolation Forest} Isolation Forest is an anomaly detection algorithm that implements a new approach compared to other models used for this purpose: rather than focussing on identifying normal points and their deviations (i.e. anomalies), Isolation Forest focuses directly on the detection of these anomalies without profiling. This can be done based on two fundamental properties of outliers, that is, they are few and different, which makes it so that they are \textit{isolated} from the other -regular- points [18]. In order to isolate anomalies, this algorithm makes use of a tree structure, which results in outliers being placed closer to the root of the tree compared to the other points [19]. In Isolation Forest, each isolation tree isolates the anomalies by randomly selecting a features and a split value between the minimum and the maximum values of that feature; the random partitioning should result in anomalies having a shorter path due to both the low number of such instances and to their inherently different characteristics leading to early partitioning. More details about the algorithm of Isolation Forest can be seen in [19]. Isolation Forest does not require labels to work, however, it is trained on the training set comprising only of legitimate transactions and used to classify the data points in the testing set.

\paragraph{K-Means}
K-Means is an unsupervised learning model used for \textit{clustering}. Clustering is the process by which from a given input, clusters or groupings are identified [20]. The process by which K-means operates can be divided into two parts: given an input comprising of a set of instances $x_1$, $x_2$, $x_3$, ..., $x_n$, and a number of clusters $K$, the algorithm places the centroids $c_1$, $c_2$, $c_3$, ...,
$c_n$ for each cluster $J$ at random locations, then:
\begin{enumerate}
\item For each point $x_n$:
\begin{enumerate}
\item Find nearest centroid $c_j$. K-means computes Euclidean distance between each point $x_n$ and centroid $c_j$. This approach is often called \textit{ minimising the inertia} of the clusters [21] and can be defined as follows: 
$$ SS_{w_i} = \sum_{n}||x_n - c_j||^2  \forall i \in (1, K) $$
Where $n$ is the number of points $x$ and $i$ is the number of centroids $c$.
\item Assign instance $x_n$ to cluster $J$. 
\end{enumerate}
\item For each cluster $J: 1,2,...K $
\begin{enumerate}
\item Compute new centroid $c_j$. This is done calculating the mean from each point $x$ to the centroid $x$ of the cluster $J$ to which is was firstly assigned. 
\end{enumerate}
\item Stop when convergence is reached, that is, there are no more changes after the iterations. 
\end{enumerate} 
For more details about K-Means, see also [21], [22]. We fit K-Means on the entire dataset specifying two clusters (for legitimate and fraudulent daily counts). The cluster containing the smallest number of instances is considered to be the cluster indicating the positive class. As with Box-Plot and LOF, only the part of the outliers in the testing set is taken into account. 

\section{Results}
%\subsection{Results of Original Time Series}

The results are presented based on three metrics: Precision, Recall and F-Measure. Precision refers to the ability of the model to be trustworthy as regards its classified positive points; that is, Precision tells us how many of the predicted frauds are actually frauds. A high Precision means that when the model classifies a point as positive it is highly likely that it is a correct classification. This metric is defined by the following equation: 
\begin{eqnarray}
\text{Precision = } \frac{\text{True Positive}}{\text{True Positive}+\text{False Positive}}
\end{eqnarray}
Recall indicates the ability of the model to detect the positive class. When a model presents a high Recall, it means that the majority of positive data points would be correctly identified. The equation for Recall is shown below. 
\begin{eqnarray}
\text{Recall = } \frac{\text{True Positive}}{\text{True Positive}+\text{False Negative}}
\end{eqnarray}
Precision and Recall indicate two opposite properties of a model, meaning that optimising one implies worsening the other. In order to gain a more comprehensive overview of the performance of the model, we can use the F-Measure metric, defined as shown in the following equation. 
\begin{eqnarray}
\text{F-Measure = } \frac{2(Precision*Recall)}{Precision+Recall}
\end{eqnarray}

These metrics are calculated for each of the 9 time series analysed and used to obtain the average as described in the previous section. The results are presented in the table \ref{table:4}.
\begin{table}[h]
\caption{Comparison of the performances for the $5$ models using the $9$ time series\vspace{2mm}}
\label{table:4}
\centering
\begin{tabular}{c|c|c|c|c|c}
\toprule
METRICS & ARIMA & BOX-PLOT & LOF & IF & K-MEANS \\
\midrule
Precision & 50\% & 43.98\% & 8.4\% & 25.01\%  & 21.82\%\\ \hline
Recall & 66.67\% & 72.22\% & 66.67\%, & 72.22\% & 83.33\%\\ \hline
F-Measure & 55.56\% & 52.22\%, & 14.04\% & 32.56\% & 28.95\%\\ 
\bottomrule
\end{tabular} 
\end{table} 

As can be noted, ARIMA presents the best result in terms of Precision and F-Measure, whereas K-Means provides the best performance in terms of Recall. The worst performing model in this setting is Local Outlier Factor, that presents a Precision and F-Measure scores equal to 8.4\% and 14.04\% respectively. It should be pointed out that LOF was designed to be effective with multidimensional datasets [16], which might explain its bad performance in this particular setting. The Box-Plot model produces the best performance amongst the benchmarks with a F-Measure of $72.22\%$ and the only one which is comparable to that of our model. The advantage of our model that it is based on the concept of modelling the normal behaviour of the customer. In addition, the forecasting by the rolling windows takes into account the dynamic changes in the spending behavior of the customer. While it can be argued that our model is overall the best one, it underperforms the Box-plot, the Isolation Forest and the K-Means in terns of Recall.\\

%\subsection{Results after Injection of Artificial Frauds}
As previously discussed, only 9 out of the 24 possible time series are retained for analysis due to the lack of frauds in the testing set. Consequently, the results that were presented are highly dependent of that particular set of data. In order to assess the robustness of the model, the time series that were originally discarded are reintegrated through the injection of one fake fraudulent transaction in the testing set. The occurrence of frauds is simulated by the addition of a varying number of counts ranging from 1 to 8 to a random date in the testing set for each time series. The range was set from 1 to 8 as it reflects the one observed in the 9 time series already discussed. It should be noted that the performance of the models highly varies depending on how many counts are added and on which day. In order to account for this randomness, this process is repeated 100 times and the average of the metrics is computed. In order to have an overview of the performances over the $24$ time series, a global average is computed and is shown in the table \ref{table:5}
%The averages are then added to the 9 original metrics to obtain global performance metrics for all 24 time series. The results can be seen in the table below. \\ 
\begin{table}[h!]
\caption{global performance for the $5$ models using the $24$ times series\vspace{2mm} }
\label{table:5}
\centering
\begin{tabular}{c|c|c|c|c|c}
\toprule
METRICS & ARIMA & BOX-PLOT & LOF & IF & K-MEANS\\ 
\midrule
Precision & 34.29\% & 28.96\% & 6.41\% & 19.94\% & 22.51\% \\ \hline
Recall & 42.03\% & 60.54\% & 69.57\%, & 64.09\% & 68.16\%\\ \hline
F-Measure & 36.19\% & 34.91\%, & 11.17\% & 24.82\% & 26.81\% \\ 
\bottomrule
\end{tabular} 
\end{table}

Despite the fact that all models under-perform after the injection of fake frauds, the ARIMA presents the best performance in terms of Precision and F-Measure, whereas the best Recall score is achieved by Local Outlier Factor. The Precision of the latter is however again the worst, which brings Box-Plot to be the only comparable model to ARIMA in this case as well. 

%As can be noted, all models appear to under-perform after the injection of fake frauds; this can be expected as the addition of data would inevitably imply adding more variety and, hence, increasing the difficulty in detecting anomalies. Despite the general worsening of performance, it is interesting to note that there is a similar trend to our original results:   ARIMA presents the best performance in terms of Precision and F-Measure, whereas the best Recall score is achieved by Local Outlier Factor. The Precision of the latter is however again the worst, which brings Box-Plot to be the only comparable model to ARIMA in this case as well. 
\section{Conclusion}

This paper addresses the problem of unsupervised approach of credit card fraud detection using the ARIMA model. The main reason on focussing on time series model comes from the lack of fraud data due to confidential issues that could represent a substantial obstacle in the development of machine learning algorithms. In this context, the goal of our approach is to model the regular spending behaviour of the customer and any discrepancies and deviations from it would be likely to be anomaly. The intuition behind this approach is centred on the assumption that the occurrence of frauds in a given day would cause the daily number of transactions to be altered in such a way that could be detected as suspicious. In the training set ARIMA model is first calibrated on the daily number of legitimate transactions in order to learn the regular spending behaviour for the customer. In the second step, the fitted model is used to predict fraud in the testing set by using the rolling windows. The criterion of flagging fraud is based on the Z-Score calculated on the prediction errors in the testing set. Our methodology is applied on the dataset that is provided by NetGuardians and is compared to $4$ anomaly detection algorithms such as K-Means, Box-Plot, Local Outlier Factor and Isolation Forest. It is shows in terms of prediction power that the ARIMA model outperforms the other models following by the Box.Plot method. Among the 4 benchmark models, the Local Outlier Factor is the worst performing model.

Our model is successful compared to the benchmarks models for two reasons:

\begin{enumerate}
\item It works better when there is a significant number of frauds happening in the same day. This is often the case, as fraudsters are known to take advantage of the time they have before the card is blocked to make several fraudulent transactions in a short time span [13]

\item It presents the best precision, i.e. it reduces the number of false positives compared to the benchmark models
\item It takes into account the dynamic spending behaviour for the customer by using the rolling windows.
\end{enumerate}

One main problem in our approach is that ARIMA model assumes that the data comes from observations that are equally spaced in time. However, this assumption does not hold in our study since the transaction times are unequally spaced. This issue will be addressed in future research by using advanced approaches such as the continuous-time autoregressive moving average (CARMA) processes.

\newpage

\section*{References}

%\medskip

\small

\noindent [1] European Central Bank, "Fifth report on card fraud", Sep, 2018.

\noindent [2] "The Nilson Report | News and Statistics for Card and Mobile Payment Executives", \textit{Nilsonreport.com}, 2020. [Online]. Available: https://nilsonreport.com/. [Accessed: 17- Feb- 2020]. 

\noindent [3] S. P. Maniraj \textit{et al}, "Credit Card Fraud Detection using Machine Learning and Data Science," \textit{International Journal of Engineering Research and Technology}, vol. 8, (9), 2019. . DOI: 10.17577/IJERTV8IS090031.

\noindent [4] D. Tripathi \textit{et al}, "Credit Dard Fraud Detection Using Local Outlier Factor," \textit{International Journal of Pure and Applied Mathematics}, vol. 118, (7), pp. 229-234, 2018.

\noindent [5] A. Dal Pozzolo \textit{et al}, "Learned lessons in credit card fraud detection from a practitioner perspective," \textit{Expert Systems with Applications}, vol. 41, (10), pp. 4915-4928, 2014. Available: http://dx.doi.org/10.1016/j.eswa.2014.02.026. DOI: 10.1016/j.eswa.2014.02.026.

\noindent [6] D. Singh, S. Vardhan and Dr. N. Agrawal, "Credit Card Fraud Detection Analysis,"   
\textit{International Research Journal of Engineering and Technology (IRJET)}, vol. 55, (11), 2018.

\noindent [7] Navanshu Khare, and Saad Yunus Sait, "Credit Card Fraud Detection Using Machine Learning Models and Collating Machine Learning Models," \textit{International Journal of Pure and Applied Mathematics}, vol. 118, (20), 2018.

\noindent [8] D. Varmedja \textit{et al}, "Credit Card Fraud Detection - Machine Learning methods," in: \textit{Proceedings of 18th International Symposium INFOTEH-JAHORINA}
, 20-22 March 2019.

\noindent [9] A. Roy \textit{et al}, "Deep learning detecting fraud in credit card transactions," in: \textit{Proceedings of the  2018 Systems and Information Engineering Design Symposium (SIEDS)}, Apr 2018, pp. 129-134

\noindent [10] R. Adhikari and R. K. Agrawal, \textit{An Introductory Study on Time Series Modeling and Forecasting}. 2013. DOI: 10.13140/2.1.2771.8084.

\noindent [11] G. E. P Box, G. M Jenkins, and G. C . Reinsel, \textit{Time Series Analysis: Forecasting and Control} (4th ed) John Wiley \& Sons, 2008.

\noindent [12] R. Azrak and G. Melard, "Exact maximum likelihood estimation for extended ARIMA models," 1993.

\noindent [13] L. Seyedhossein and M. R. Hashemi, "Mining information from credit card time series for timelier fraud detection," \textit{International Journal of Information and Communication Technology}, vol 2 (3), Nov 2010, pp. 619-624.

\noindent [14] S. Ounacer \textit{et al}, "Using Isolation Forest in anomaly detection: the case of credit card transactions," \textit{Periodicals of Engineering and Natural Sciences (PEN)}, vol. 6, (2), pp. 394, 2018. . DOI: 10.21533/pen.v6i2.533.

\noindent [15] D. F. Williamson, "The Box Plot: A Simple Visual Method to Interpret Data," \textit{Annals of Internal Medicine}, vol. 110, (11), pp. 916, 1989. DOI: 10.7326/0003-4819-110-11-916.

\noindent [16] M. M. Breunig \textit{et al}, "LOF," \textit{ACM SIGMOD Record}, vol. 29, (2), pp. 93-104, 2000. DOI: 10.1145/335191.335388.

\noindent [17] "sklearn.neighbors.LocalOutlierFactor-scikit-learn 0.22.1 documentation", Scikit-learn.org, 2020. [Online]. [Accessed: 20- Feb- 2020].

\noindent [18] H. John and S. Naaz, "Credit Card Fraud Detection using Local Outlier Factor and Isolation Forest," \textit{International Journal of Computer Sciences and Engineering}, vol. 7, (4), pp. 1060-1064, 2019.
DOI: 10.26438/ijcse/v7i4.10601064.

\noindent [19] F. T. Liu, Kai Ming Ting and Zhi-Hua Zhou, "Isolation forest," in: \textit{Proceedings of the 8th IEEE International Conference on Data Mining}, Pisa, Italy, 2008, pp. 413-422.

\noindent [20] M. Kubat, \textit{An Introduction to Machine Learning}, 2015. DOI: 10.1007/978-3-319-20010-1.

\noindent [21] P. I. Dalatu, " Time Complexity of K-Means and K-Medians Clustering Algorithms in Outliers Detection," \textit{Global Journal of Pure and Applied Mathematics}, vol. 12, (5), pp. 4405-4418, 2016.

\noindent [22] G. Bonaccorso, \textit{Machine Learning Algorithms: Popular Algorithms for Data Science and Machine Learning}, 2018.

\end{document}